\documentclass{article}

    \usepackage[preprint]{neurips_2022}

\usepackage[utf8]{inputenc} 
\usepackage[T1]{fontenc}    
\usepackage{hyperref}       
\usepackage{url}            
\usepackage{booktabs}       
\usepackage{amsfonts}       
\usepackage{nicefrac}       
\usepackage{microtype}      
\usepackage{xcolor}         
\usepackage{amsmath}

\usepackage{subcaption, graphicx, wrapfig, booktabs, lipsum}
\usepackage{enumitem}
\usepackage{natbib}
\setcitestyle{square,numbers,comma}

\def\pr{\textsf{P}}
\def\ep{\textsf{E}}
\def\var{\textsf{Var}}
\def\cov{\textsf{Cov}}

\usepackage{algorithm}
\usepackage{algorithmic}

\newtheorem{theorem}{Theorem}[section]
\newtheorem{remark}[theorem]{Remark}

\title{New graph-based multi-sample tests for high-dimensional and non-Euclidean data}

\author{%
  Hoseung Song \\
  Public Health Sciences Division \\
  Fred Hutchinson Cancer Research Center\\
  Seattle, Washington 98109 \\
  \texttt{hsong3@fredhutch.org} \\
  \And
  Hao Chen \\
  Department of Statistics \\
  University of California, Davis \\
  Davis, California 95616 \\
  \texttt{hxchen@ucdavis.edu} \\
}

\begin{document}

\maketitle

\begin{abstract}
  Testing the equality in distributions of multiple samples is a common task in many fields. However, this problem for high-dimensional or non-Euclidean data has not been well explored. In this paper, we propose new nonparametric tests based on a similarity graph constructed on the pooled observations from multiple samples, and make use of both within-sample edges and between-sample edges, a straightforward but yet not explored idea. The new tests exhibit substantial power improvements over existing tests for a wide range of alternatives. We also study the asymptotic distributions of the test statistics, offering easy off-the-shelf tools for large datasets. The new tests are illustrated through an analysis of the age image dataset.
\end{abstract}


\section{Introduction} \label{sec:intro}

Testing the equality of underlying distributions is a classical problem. As we entering the big data era, many problems involve the test on high-dimensional data or even non-Euclidean data and testing the homogeneity in distributions of more than two independent samples is gaining attention in many applied research fields. Formally speaking, given $K$ independent samples $Y_{11}, Y_{12}, \ldots, Y_{1n_{1}} \stackrel{iid}{\sim} F_{1}$, $Y_{21}, Y_{22}, \ldots, Y_{2n_{2}} \stackrel{iid}{\sim} F_{2}$, $\ldots$, $Y_{K1}, Y_{K2}, \ldots, Y_{Kn_{K}} \stackrel{iid}{\sim} F_{K}$, we concern the following hypothese testing:
\begin{equation} \label{problem}
H_{0}: \ F_{1} = F_{2} = \cdots = F_{K} \ \ \textrm{versus} \ \ H_{1} : \exists \ i,j \ \textrm{s.t} \ F_{i} \ne F_{j}.
\end{equation}
Here, $Y_{ij}$'s can be a high-dimensional vector or a non-Euclidean data object. Following are some motivating examples.
\begin{itemize}
	\item Identification of genetic pathways : A simultaneous analysis of a genetic pathway might provide clear insight into cause of the phenotypic changes \cite{glaab2012enrichnet,luo2009gage}. For example, the identification of important genetic pathways that drive the cancer progression may provide insights into the understanding of molecular mechanism of cancer progression \cite{zhang2020graph}. Here, each observation is gene expression and the $K$-samples are development stages of cancer.
	\item Survey data: One goal of survey is to detect any difference in the pattern of answers given by different groups of respondents \cite{nettleton2001testing}. In this case, multi-sample comparison would be useful since it is computationally more efficient than the pairwise comparison for large number of groups of respondents. Here, each observation is the survey result, e.g., binary/multiple-choice responses of questions, and the $K$-samples are groups of respondents.
	\item Pricing in insurance data:  Dealing with insurance data, it is often of interest to compare several portfolios or groups and this is particularly useful in pricing for risk pooling or price segmentation \cite{shi2016multilevel}. Here, the database presents a summary of claims and each observation consists of general information about claimant, such as medical charges or expenses \cite{bakam2021k,grazier2004group}. Then, each observation is claim information and the $K$-samples can be specific groups or different time intervals.
\end{itemize}

In this paper, we focus on $K>2$. The multi-sample test has been extensively stuided for univariate data \cite{kruskal1952nonparametric, lemeshko2018some, zhang2007k}. Recently, there are many advances for dealing with high-dimensional data as well. For example, \cite{cai2017hypothesis} proposed an empirical likelihood method through a density ratio model. \cite{chen2020new} proposed a distribution-free multi-sample test based on the analysis of kernel density functional estimation. However, these methods mainly focus on differences in the mean. There are other methods using MANOVA \cite{rizzo2010disco}, the maximum mean discrepancy \cite{kim2019comparing}, or a spectral graph partitioning \cite{mukhopadhyay2020nonparametric}; however these methods are computationally extensive for large datasets.

Recently, graph-based two-sample tests attracted attention due to their flexibilities \cite{chen2017weighted, chen2017new, friedman1979multivariate, henze1988multivariate, rosenbaum2005exact,schilling1986multivariate}. This line of work leads to some promising generalizations for $K$-sample comparsion. For example, \cite{nettleton2001testing} considered the generalization of the methods proposed by \cite{schilling1986multivariate} and \cite{ henze1988multivariate}, which utilize the nearest neighbor graph. \cite{petrie2016graph} generalized the method proposed by \cite{friedman1979multivariate} for multi-sample problems, which counts the number of edges between samples in the minimum spanning tree (MST), a spanning tree that connects all observations with the sum of distances of the edges in the tree minimized. \cite{agarwal2019distribution} generalized the method proposed by \cite{rosenbaum2005exact} and proposed distribution-free multi-sample tests based on the minimum non-bipartite matching of the pooled sampled.

All these graph-based tests are robust and computationally efficient, and perform well for high-dimensional/non-Euclidean data. However, although they were proposed for general alternatives, they are not powerful for some common types of alternatives when the dimension is high.

To address this, we propose new graph-based tests that work for a wide range of alternatives. Given the similarity graph on $K$-samples, we take a different approach from the existing graph-based multi-sample tests in that we utilize both within-sample edges and between-sample edges so that the test statistic contains as much information as possible. We investigate a few combinations and study the asymptotic distributions of the new tests to make them computationally efficient for large datasets. Simulation experiments show that the new tests exhibit high power in both synthetic and real world data.


\section{New test statistics} \label{sec:new}

\begin{figure}
	\centering
	\includegraphics[width=3.5in]{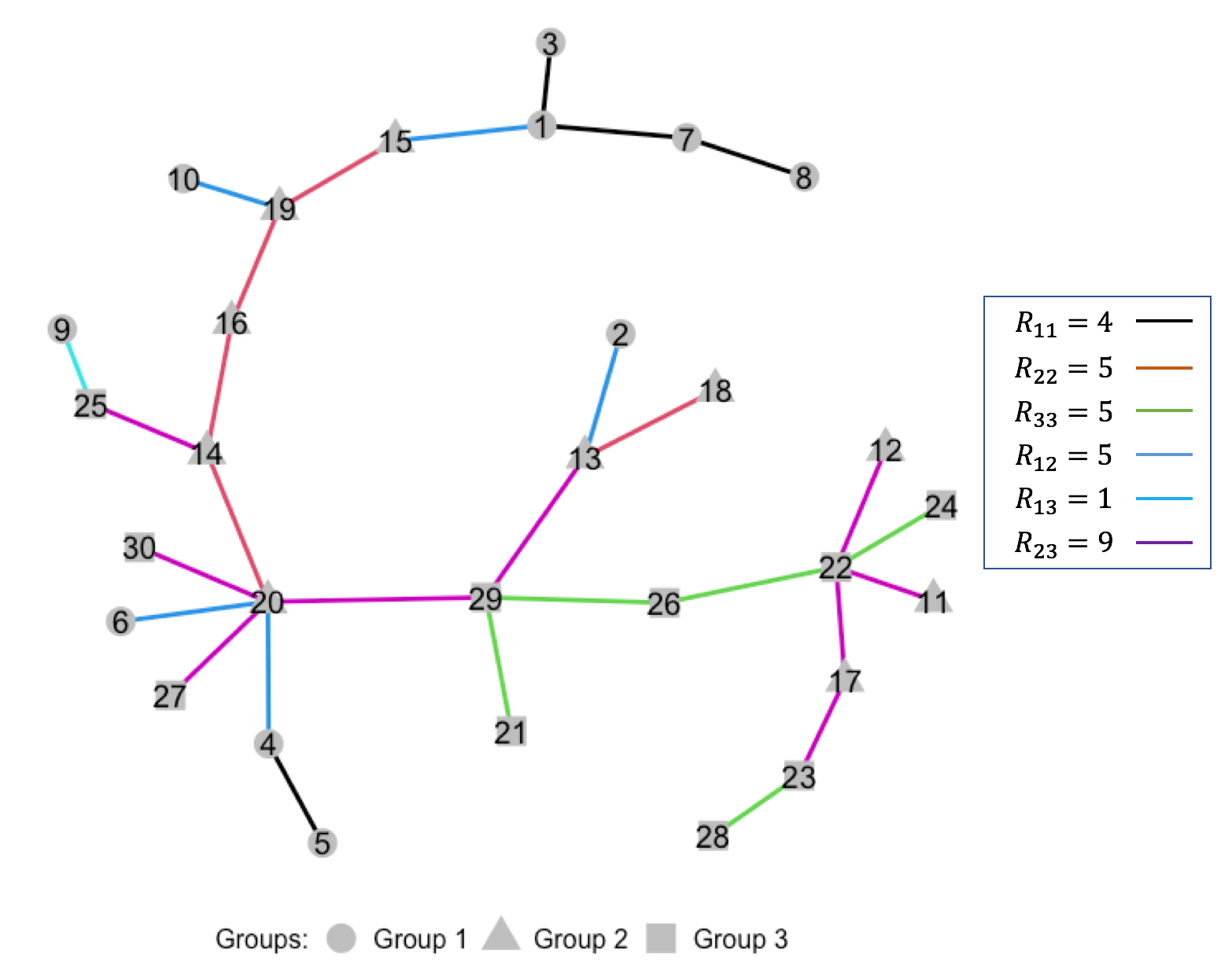}
	\caption{Illustration of edges in the MST on pooled observations from $N_{d}(0.3(i-1)\textbf{1}_{d}, I_{d})$ where $K=3$, $d=50$, $n_{i}=10$, and $\textbf{1}_{d}$ is $d$ dimensional vector of ones $(i=1,2,3)$.} 
	\label{fig:graph}
\end{figure}

Let $N = \sum_{i=1}^{K}n_{i}$ be the total sample size. Define $G$ as the similarity graph, e.g., the MST, on the pooled observations $\{Y_{i1}, \ldots, Y_{in_{i}}\}$ for $i = 1,\ldots,K$. We work under the permutation null distribution, which places $(n_{1}!\cdots n_{K}!)/N!$ probability on each of the $N! /(n_{1}!\cdots n_{K}!)$ choices of $n_{i}$ out of the total $N$ observations for $i = 1, \ldots, K$. With no further specification, $\pr$, $\ep$, $\var$, and $\cov$ denote the probability, expectation, variance, and covaraince, repectively, under the permutation null distribution.
We define $R_{ij}$ as the number of edges in $G$ with one endpoint in sample $i$, $\{Y_{i1}, \ldots, Y_{in_{i}}\}$, and the other endpoint in sample $j$, $\{Y_{j1}, \ldots, Y_{jn_{j}}\}$. Figure \ref{fig:graph} illustrates $R_{ij}$'s on the MST constructed on the pool of three samples.

Unlike two-sample comparisons, many types of interactions could exist under the $K$-sample comparison. The existing graph-based multi-sample tests only consider information in either within-sample edges ($R_{ii}$'s) or between-sample edges ($R_{ij}$'s, $i\ne j$). For example, the test statistics in \cite{agarwal2019distribution}, \cite{nettleton2001testing}, and \cite{petrie2016graph} only consider the between-sample edges.  Moreover, the test statistics in \cite{nettleton2001testing}, and \cite{petrie2016graph}  and a test statistic in \cite{agarwal2019distribution} use the sum of all between-sample edges, which is equivalent to using the sum of all within-sample edges.    In the two-sample setting,  $R_{11}+R_{12}+R_{22}$ equals to a constant (the number of edges in $G$).  Thus, only using the two within-sample edges is sufficient \citep{chen2017new}. In the $K$-sample setting, only using within-sample edges or only using between-sample edges can cause information loss. Thus, we explore to use them together.

Let $\textbf{R}^{W}$ be the vector of  $K$ $R_{ii}$'s, $i=1,\ldots,K$, and $\textbf{R}^{B}$ be the vector of  $K(K-1)/2$ $R_{ij}$'s, $1\le i < j \le K$. We consider two Mahalanobis-like quantities:
\begin{align}
S^{W} &= \left(\textbf{R}^{W} - \ep\big(\textbf{R}^{W}\big)\right)^{T}\Sigma_{W}^{-1}\left(\textbf{R}^{W} - \ep\big(\textbf{R}^{W}\big)\right), \label{sw} \\
S^{B} &= \left(\textbf{R}^{B} - \ep\big(\textbf{R}^{B}\big)\right)^{T}\Sigma_{B}^{-1}\left(\textbf{R}^{B} - \ep\big(\textbf{R}^{B}\big)\right), \label{sb}
\end{align}
where $\Sigma_{W} = \var\big(\textbf{R}^{W}\big)$ and $\Sigma_{B} = \var\big(\textbf{R}^{B}\big)$. For location alternatives, observations from the same distribution would be preferentially closer to each other than observations from different distributions. Hence, when the null hypothesis (\ref{problem}) is not true, $\textbf{R}^{W}$ tends to be larger than its null expectation, making $S^{W}$ large. On the other hand, $\textbf{R}^{B}$ tends to be smaller than its null expectation, leading to large $S^{B}$ as well. For scale alternatives, most observation from the distribution having a larger variance tend to be closer to observations from the distribution with a smaller variance due to the curse of dimensionality (see details in \cite{chen2017new}). This makes some $R_{ii}$'s smaller or larger than its null expecation, which leads to large $S^{W}$. For more complicated alternatives, there could be many possible scenarios.  Using both within-sample and between-sample edges could catch more types of alternatives than using one of them only.  Based on $S^{W}$ and $S^{B}$, one way to combine the information is to add them:
\begin{equation}
S = S^{W} + S^{B}.
\end{equation}
We also define a test statistic based on all linearly independent $R_{ij}$'s. Let $\textbf{R}^{A}$ be $\textbf{R}^{A} = \left(R_{11}, R_{22}, \ldots, R_{KK}, R_{12}, R_{13}, \ldots, R_{(K-2)K}\right)^{T}$. Notice that $R_{(K-1)K}$ is left out. We then define another test statistic as
\begin{align}
S^{A} &= \left(\textbf{R}^{A} - \ep\big(\textbf{R}^{A}\big)\right)^{T}\Sigma_{A}^{-1}\left(\textbf{R}^{A} - \ep\big(\textbf{R}^{A}\big)\right), \Sigma_{A} = \var\big(\textbf{R}^{A}\big).
\end{align}

\begin{figure}[h]
	\centering
	\includegraphics[width=\columnwidth]{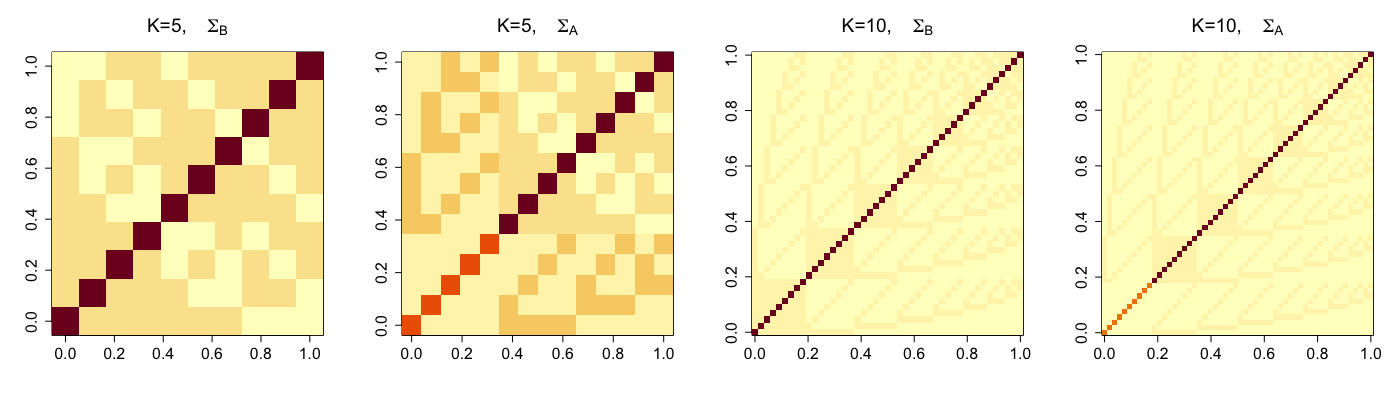}
	\caption{Plots of $\Sigma_{B}$ and $\Sigma_{A}$ for Gaussian data when $K=5\&10$, $d=50$, and $n_{i}=30$.} 
	\label{fig:inverse}
\end{figure}

It can be shown that $\Sigma_W$ is always invertible \citep{zhang2020graph}.  For $\Sigma_B$ and $\Sigma_A$, it is difficult to check their invertibility theoretically.  We check them numerically by simulating data from the following distributions: the multivariate Gaussian data $N_{d}(\textbf{0}_{d},\Sigma)$, multivariate log-normal data $\exp(N_{d}(\textbf{0}_{d},\Sigma))$, multivariate $t$-distributed data $t_{20}(\textbf{0}_{d},\Sigma)$, and chi-square data $\Sigma^{1/2}u$, where $u$ is length-$d$ vectors with each component i.i.d. from the $\chi_{3}^2$ distribution,  $\textbf{0}_{d}$ is $d$ dimensional vector of zeros, and $\Sigma_{(i,j)} = 0.4^{|i-j|}$. Here, we set $d$ and $n_{i}$ to be 50 and 30, respectively. When we check 1,000 datasets for each distribution and $K = 3,\ldots,10$, the covariance matrices are invertible in all cases.  Figure \ref{fig:inverse} plots a typical result of $\Sigma_B$ and $\Sigma_A$ for Gaussian data when $K=5$ and 10.  We see that the diagonal elements dominate the other elements in both $\Sigma_B$ and $\Sigma_A$, making them nonsingular.   In practice, one can  check the covariance matrix first to see whether it is invertible before applying the method.  If it is not, the generalized inverse can be used instead.

We pool observations from all $K$ samples and index them by $1, 2, \dots, N$.   Let $g_{t}^{(i)} = 1$ if observation $t$ is from sample $i$ and 0 otherwise for $t = 1,\ldots,N$ and $i=1,\ldots,K$. Let $I_{x}$ be the indicator function. For an edge $e = (u,v) \in G$, $R_{ij}$ can be rewritten as
\begin{equation*}
R_{ij} = \sum_{e\in G}I_{g_{u}^{(i)} = g_{v}^{(j)}=1}.
\end{equation*}
Let $G_{t}$ be the subgraph of $G$ containing all edges that connect to node $t$ and $|G_{t}|$ be the number of edges in $G_{t}$ or the degree of node $t$ in $G$. Let $|G|$ be the total number of edges in $G$. Theorem \ref{thm:stat} provides the analytic expressions of the expectation and the variance of $R_{ij}$ under the permutation null.
\begin{theorem} \label{thm:stat}
	Under the permutation null, we have
	{\small
	\begin{align*}
	\ep\left(R_{ii}\right) &= |G|\frac{n_{i}(n_{i}-1)}{N(N-1)}, \ \ \ \ep\left(R_{ij}\right) = |G|\frac{2n_{i}n_{j}}{N(N-1)}, \\
	\var\left(R_{ii}\right) &= |G|\frac{n_{i}(n_{i}-1)}{N(N-1)} + \left(\sum_{t=1}^{N}|G_{t}|^2-2|G|\right)\frac{n_{i}(n_{i}-1)(n_{i}-2)}{N(N-1)(N-2)} \\
	& \ \ \ \ \ \ \ \  + \left(|G|^2-\sum_{t=1}^{N}|G_{t}|^2+|G|\right)\frac{n_{i}(n_{i}-1)(n_{i}-2)(n_{i}-3)}{N(N-1)(N-2)(N-3)} - \ep\left(R_{ii}\right)^2, \\
	\var\left(R_{ij}\right) &= |G|\frac{2n_{i}n_{j}}{N(N-1)} + \left(\sum_{t=1}^{N}|G_{t}|^2-2|G|\right)\frac{n_{i}n_{j}\left(n_{i}+n_{j}-2\right)}{N(N-1)(N-2)} \\
	&\ \ \ \ \ \ \ \  + \left(|G|^2-\sum_{t=1}^{N}|G_{t}|^2+|G|\right)\frac{4n_{i}n_{j}(n_{i}-1)(n_{j}-1)}{N(N-1)(N-2)(N-3)} - \ep\left(R_{ij}\right)^2, \\
	\cov\left(R_{ii},R_{jj}\right) &= \left(|G|^2-\sum_{t=1}^{N}|G_{t}|^2+|G|\right)\frac{n_{i}n_{j}(n_{i}-1)(n_{j}-1)}{N(N-1)(N-2)(N-3)}  - \ep\left(R_{ii}\right)\ep\left(R_{jj}\right), \\
	\cov\left(R_{ii},R_{ij}\right) &= \left(\sum_{t=1}^{N}|G_{t}|^2-2|G|\right)\frac{n_{i}n_{j}(n_{i}-1)}{N(N-1)(N-2)} \\
	&\ \ \ \ \ \ \ \  + 2\left(|G|^2-\sum_{t=1}^{N}|G_{t}|^2+|G|\right)\frac{n_{i}n_{j}(n_{i}-1)(n_{i}-2)}{N(N-1)(N-2)(N-3)} - \ep\left(R_{ii}\right)\ep\left(R_{ij}\right), \\
	\cov\left(R_{ii},R_{jk}\right) &=   \left(|G|^2-\sum_{t=1}^{N}|G_{t}|^2+|G|\right)\frac{2n_{i}n_{j}n_{k}(n_{i}-1)}{N(N-1)(N-2)(N-3)}   - \ep\left(R_{ii}\right)\ep\left(R_{jk}\right), \\
	\cov\left(R_{ij},R_{ik}\right) &= \left(\sum_{t=1}^{N}|G_{t}|^2-2|G|\right)\frac{n_{i}n_{j}n_{k}}{N(N-1)(N-2)} \\
	& \ \ \ \ \ \ \ \     + \left(|G|^2-\sum_{t=1}^{N}|G_{t}|^2+|G|\right)\frac{4n_{i}n_{j}n_{k}(n_{i}-1)}{N(N-1)(N-2)(N-3)}    - \ep\left(R_{ij}\right)\ep\left(R_{ik}\right)\\
	\cov\left(R_{ij},R_{kl}\right) &= \left(|G|^2-\sum_{t=1}^{N}|G_{t}|^2+|G|\right)\frac{4n_{i}n_{j}n_{k}n_{l}}{N(N-1)(N-2)(N-3)}   - \ep\left(R_{ij}\right)\ep\left(R_{kl}\right),
	\end{align*}}
	for $1 \le  i\ne j\ne k\ne l \le K$.
\end{theorem}
This theorem can be proved by combinatorial analysis and details are in Appendix A.


\section{Asymptotics and fast tests} \label{sec:asymp}

Given the new test statistics, the next step is to determine how large the test statistics are to provide enough evidence to reject the null hypothesis. In our framework, the cutoffs for the new tests can be obtained from the permutation null distribution. However, this approach is time-consuming when the sample size is  large. Hence, we study the asymptotic distribution of the test statistics under the usual limiting regime: as $N\rightarrow\infty$,
\begin{align*}
\left(n_{1}/N,\ldots,n_{K}/N\right) \rightarrow \left(\lambda_{1},\ldots,\lambda_{K}\right) \in (0,1)^{K},
\end{align*}
where $\sum_{i=1}^{K}\lambda_{i} = 1$. For the similarity graph $G$ and edges $e\in G$, we define
\begin{align*}
A_{e} &= \{e\} \cup \{e'\in G: e' \ \textrm{and} \ e \ \textrm{share a node} \}, \\
B_{e} &= A_{e} \cup \{e'' \in G: e'' \ \textrm{and} \ e'\in A_{e} \ \textrm{share a node} \}.
\end{align*}
In other words, $A_{e}$ is a set of edges in $G$ connecting to an edge $e$ and $B_{e}$ is a set of edges in $G$ that connect to any edges in $A_{e}$.
\begin{theorem} \label{thm:asymp}
	If $|G| \sim O(N)$, $\sum_{t=1}^{N}|G_{t}|^2 - 4|G|^2/N \sim O(N)$, and $\sum_{e\in G}|A_{e}|B_{e}| \sim o(N^{1.5})$, in the usual limiting regime, under the permutation null, 
	\begin{align*}
	S^{W} \rightarrow \chi_{K}^2, \ \ S^{B} \rightarrow \chi_{b}^2, \ \ S^{A} \rightarrow \chi_{a}^2,
	\end{align*} 
	where $b = rank(\Sigma_{B})$ and $a = rank(\Sigma_{A})$.
\end{theorem}
The proof is provided in Appendix B.

\begin{remark}
	The proof of Theorem \ref{thm:asymp} extends the method in \cite{chen2017new}. The asymptotic distribution of $S^{W}$ is also studied in \cite{zhang2020graph}, but in this paper we take a different approach in that we first prove the asymptotic distribution of $S^{A}$ and then utilize the property of the multivariate normal distribution and quadratic forms in singular normal variables studied in \cite{styan1970notes}.
\end{remark}

\begin{remark}
	Conditions in Theorem \ref{thm:asymp} prevent both the size and number of clusters having a large degree in $G$ (so-called hub). It was shown that all conditions in Theorem \ref{thm:asymp} are satisfied when the graph is the $k$-MST, $k=O(1)$, the union of the 1st, $\ldots$, $k$-th MSTs where the $k$-th MST is the MST which does not contain any edges in the 1st, $\ldots$, $(k-1)$-th MSTs, based on the Euclidean distance for multivariate data \cite{chen2017new}.
\end{remark}

\begin{figure}[h]
	\centering
	\includegraphics[width=\columnwidth]{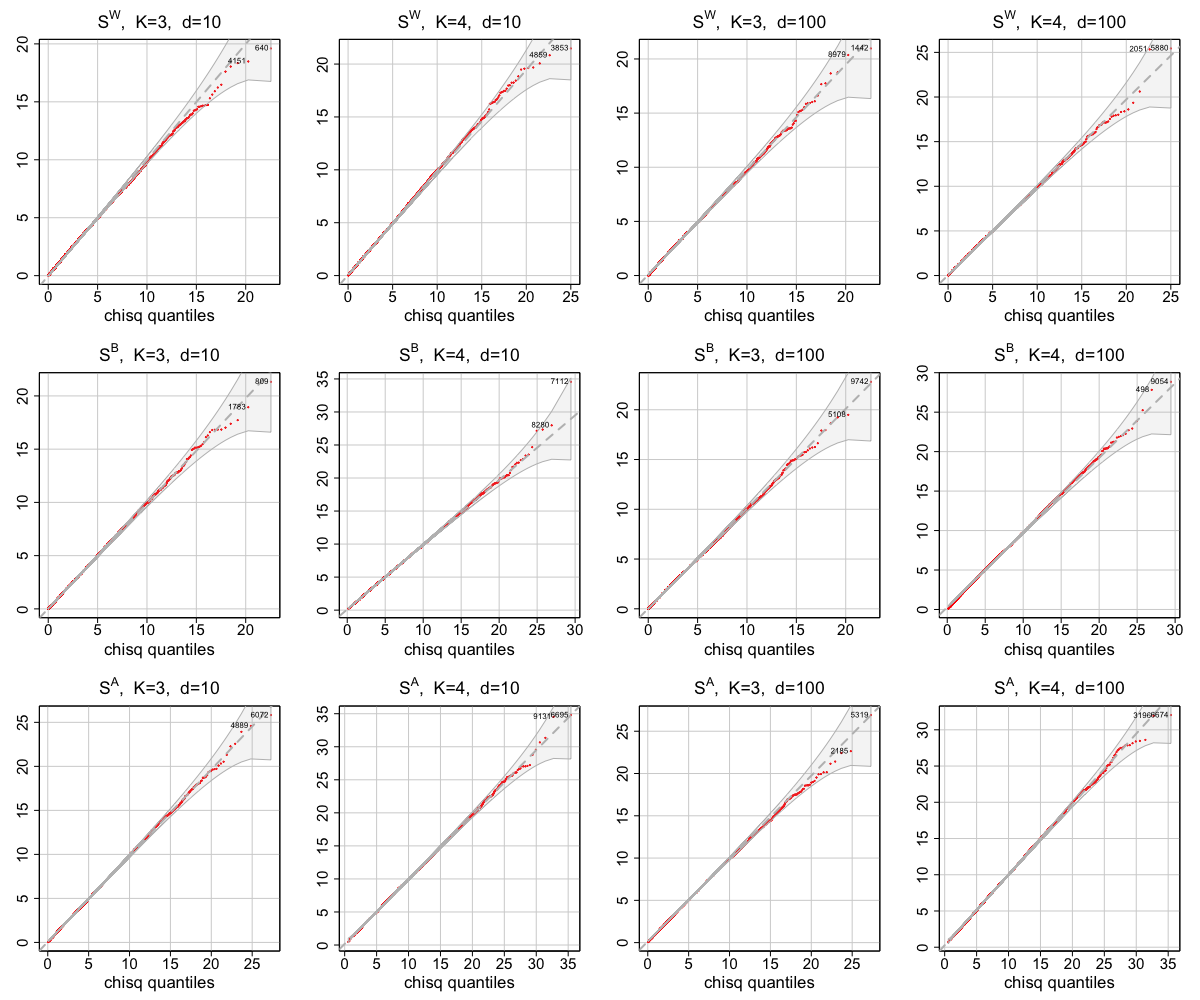}
	\caption{Quantile-quantile plots of $S^{W}$, $S^{B}$, and $S^{A}$ for Gaussian data $N_{d}(\textbf{0}_{d}, I_{d})$ when $n_{i}=50$.} 
	\label{fig:qq}
\end{figure}

The rank of $\Sigma_{B}$ and $\Sigma_{A}$ can be calculated from the function \texttt{rankMatrix}() in the \texttt{R} package \texttt{Matrix}. Figure \ref{fig:qq} shows the chi-square quantile-quantile plots of $S^{W}$, $S^{B}$, and $S^{A}$ under different choices of $K$ and $d$ when $n_{i}=50$. We see that the asymptotic distributions of $S^{W}$, $S^{B}$, and $S^{A}$ can be well approximated by the chi-square distribution.

\begin{remark}
	According to \cite{ferrari2019note}, under the conditions in Theorem \ref{thm:asymp} and the permutation null, in the usual limiting regime, the asymptotic distribution of $S$ can be obtained as
	\begin{equation*}
	S \rightarrow Gamma\left(\frac{K^2(K+1)^2}{4K(K+1)+8\rho}, \frac{4\rho+2K(K+1)}{K(K+1)}\right), 
	\end{equation*}
	where $\rho = \lim_{N\rightarrow\infty}\cov\left(S^{W}, S^{B}\right)$. Despite the asymptotic distribution of $S$, in order to apply $S$ in practice, we need to use the finite sample version of the covariance between $S^{W}$ and $S^{B}$. However, it requires fourth moments and the computation is very complicated. To make use of the asymptotic results of $S^{W}$ and $S^{B}$ and combine the advantages of the two statistics, we adopt the Bonferroni correction on $S^{W}$ and $S^{B}$. Let $p_{W}$ and $p_{B}$ be the approximated $p$-values of the tests based on $S^{W}$ and $S^{B}$, respectively. Then, the proposed test rejects the null hypothesis if $p = 2\min\left(p_{W},p_{B}\right)$ is less than the significance level of the test. The fast test, denoted by $SS$, is summarized in Algorithm \ref{algo}. Hence, as long as $S^{W}$ and $S^{B}$ are computed, the $p$-value of the new test based on $S$ can be obtained instantly.
\end{remark}

\begin{algorithm}[h]
	\caption{Fast test $SS$} \label{algo}
	\begin{algorithmic}[1]
		\REQUIRE Observations $\{Y_{ij}\}_{i=1,\dots,K; j=1,\dots, n_i}$ and the significance level $\alpha$.
		\ENSURE Reject the null hypothesis $H_{0}$ (\ref{problem}) if $p$-value $p \le \alpha$.
		\STATE Construct the similarity graph $G$, such as MST, on pooled observations.
		\STATE Compute $S^{W}$ and $S^{B}$ by ($\ref{sw}$) and ($\ref{sb}$) based on Theorem \ref{thm:stat}.
		\STATE Calculate $p$-values of $S^{W}$ and $S^{B}$ ($p_{W}$ and $p_{B}$) by the chi-square distributions in Theorem \ref{thm:asymp}.
		\STATE Obtain $p$-value $p = 2\min\left(p_{W},p_{B}\right)$.
	\end{algorithmic}
\end{algorithm}

\cite{henze1999multivariate} showed that the graph-based two-sample test using the MST is consistent against all alternatives. This Henze-Penrose divergence between probability measures \cite{ali1966general, henze1999multivariate} provides one direction to understand the consistency of graph-based two-sample tests, such as \cite{friedman1979multivariate}, \cite{henze1988multivariate}, and \cite{rosenbaum2005exact}.  An extension of these arguments can be adapted to show the consistency of the new test statistics against all alternatives in the multivariate setting. 
\begin{theorem} \label{thm:consistency}
	In the usual limiting regime, if the graph is the $k$-MST, $k = O(1)$,  based on the Euclidean distance for multivariate data, the test with rejection $\{S \ge s\}$ is universally consistent. If $\Sigma_{A}$ is invertible, the test with rejection $\{S^{A} \ge s_{A}\}$ is also universally consistent.
\end{theorem}
The proof of this theorem is in Appendix C.


\section{Numerical Experiments} \label{sec:numerical}

In this section, we examine the performance of the new tests under various settings. To this end, we follow the simulation setup in \cite{agarwal2019distribution} and compare the new tests with other graph-based tests: the multi-sample Friedman-Rafsky test (FR) proposed by \cite{petrie2016graph} and the MCM and MMCM tests proposed by \cite{agarwal2019distribution}, which can be implemented by an $\texttt{R}$ package $\texttt{multicross}$. Here, we denote the tests based on $S$ and $S^{A}$ by $S$ and $A$, respectively, and the Bonferrnoi test on $S^{W}$ and $S^{B}$ by $SS$.

Some previous works \citep{chen2017weighted, chen2017new,friedman1979multivariate} suggested to use the $k$-MST as $G$ to improve the power of the tests. Here, we use the 5-MST for S, SS, A, and FR. In all the following experiments, the significance level is set to be 0.05 and the empirical power is estimated by 1,000 iterations.

We consider the following scenarios (more simulation results can be found in Appendix D):
\begin{itemize}
	\item Location (S1): $i$-th distribution is $N_{d}\left(\mu(i-1)\textbf{1}_{d},I_{d}\right)$ $(1\le i\le K)$ where $n_{i}=50$, $\mu=0.14$ $(K=3)$, $\mu=0.1$ $(K=4)$, $\mu=0.07$ $(K=5)$, $d\in \{50, 100, 200, 300\}$.
	\item Scale (S2): $i$-th distribution is $N_{d}\left(\textbf{0}_{d},\{1+\sigma^2(i-1)\}I_{d}\right)$ $(1\le i\le K)$  where $n_{i}=50$, $\sigma^2=0.08$ $(K=3)$, $\sigma^2=0.05$ $(K=4)$, $\sigma^2=0.07$ $(K=5)$, $d\in \{50, 100, 200, 300\}$.
	\item Covariance (S3): $i$-th distribution is $N_{d}(\textbf{0}_{d},\Sigma^{(i)})$ where $K=3$, $n_{i}=50$, $d=100$, $\Sigma_{uv}^{(i)} = \rho_{i}^{|u-v|}$, and $\rho_{i} = 0.1+\sigma^2(i-1)$ ($\sigma^2\in \{0.15, 0.2, 0.25, 0.3\}$).
	\item Kurtosis (S4): Observations in each coordinate are from independent $t$ distributions and they are standardized. Here, $K=3$, $n_{i}=50$, $d=100$. $i$-th distribution has the degree of freedom $\textrm{df}=\nu_{i}$ where $\nu_{i}=2+(i-1)\Delta$ ($\Delta\in \{0.1, 0.2, 0.3, 0.4\}$, $i=1,2,3$).
	\item Skewness and kurtosis (S5): Observations in each coordinate are from independent chi-square distributions and they are standardized. Here, $K=3$, $n_{i}=50$, $d=100$. $i$-th distribution has $\textrm{df}=\nu_{i}$ where $\nu_{i}=1+(i-1)\Delta$ ($\Delta\in \{1,2,3,4\}$, $i=1,2,3$).
	\item Multivariate log-normal data (S6): $i$-th distribution is $\exp(N_{d}(0.04(i-1)\textbf{0}_{d},\Sigma))$ for location alternatives and $\exp(N_{d}(\textbf{0}_{d},(1+0.05(i-1))\Sigma))$ for scale alternatives $(1\le i \le K)$, where $K\in \{4,6,8,10\}$, $d=200$, $\Sigma_{uv} = 0.4^{|u-v|}$.
	\item Multivariate $t$-distributed data (S7): $i$-th distribution is $t_{20}(0.04(i-1)\textbf{0}_{d},\Sigma)$ for location alternatives and $t_{20}(\textbf{0}_{d},(1+0.1(i-1))\Sigma)$ for scale alternatives $(1\le i \le K)$, where $K\in \{4,6,8,10\}$, $d=200$, $\Sigma_{uv} = 0.4^{|u-v|}$.
\end{itemize}

\begin{figure}[htp!]
	\centering
	\includegraphics[width=\columnwidth]{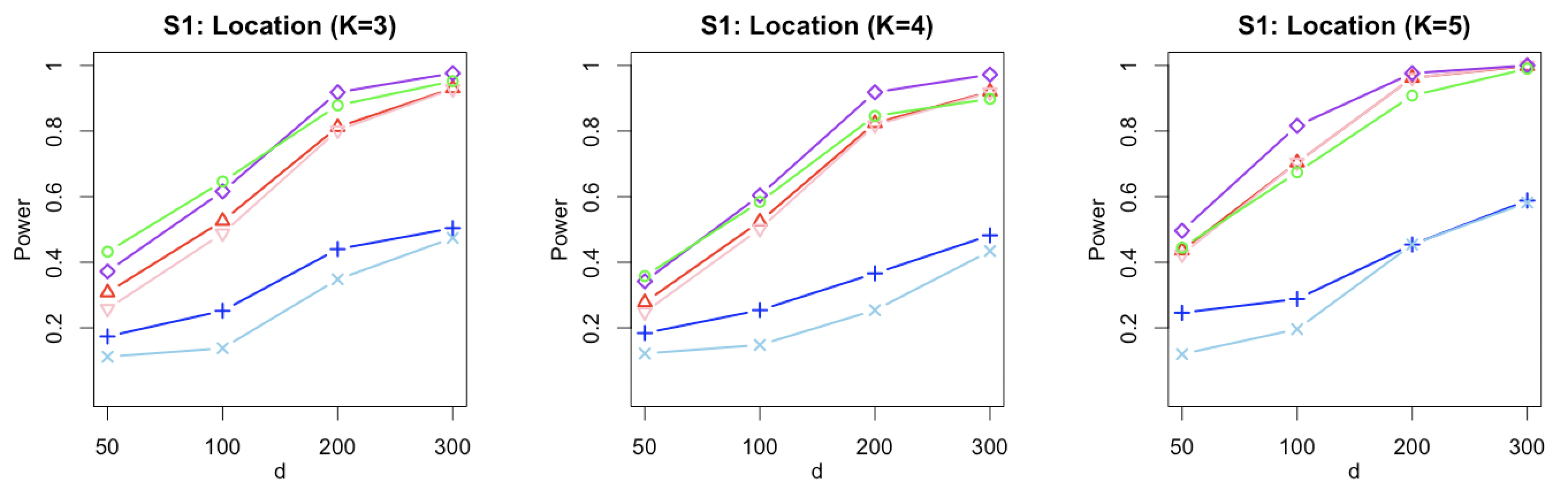}
	\includegraphics[width=\columnwidth]{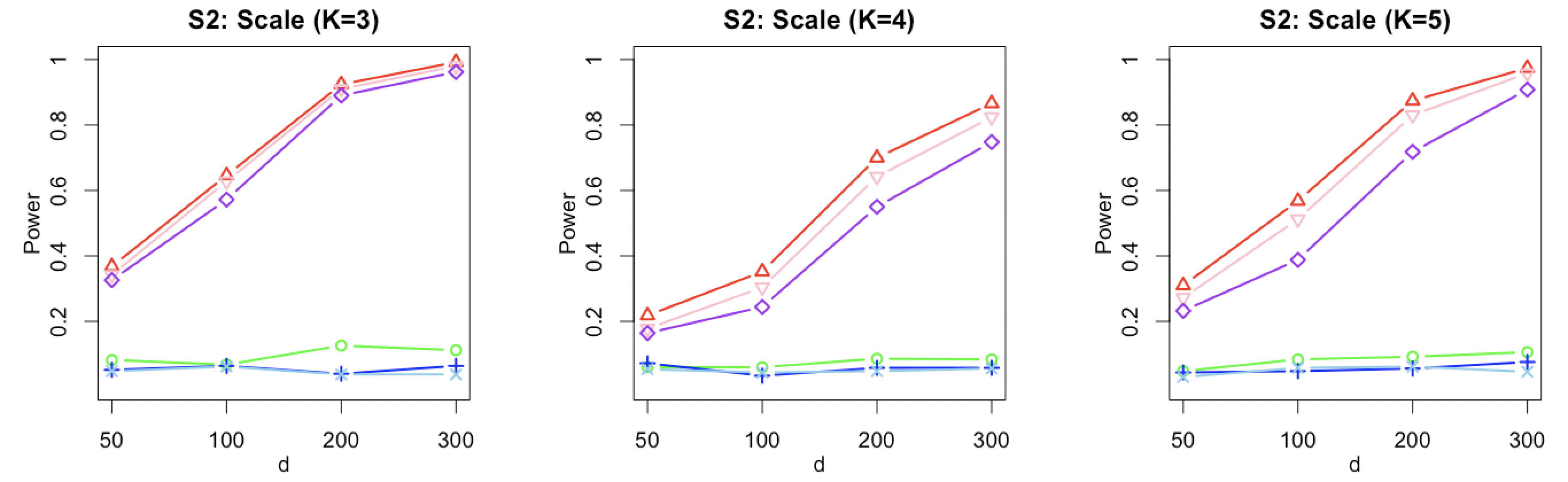}
	\includegraphics[width=\columnwidth]{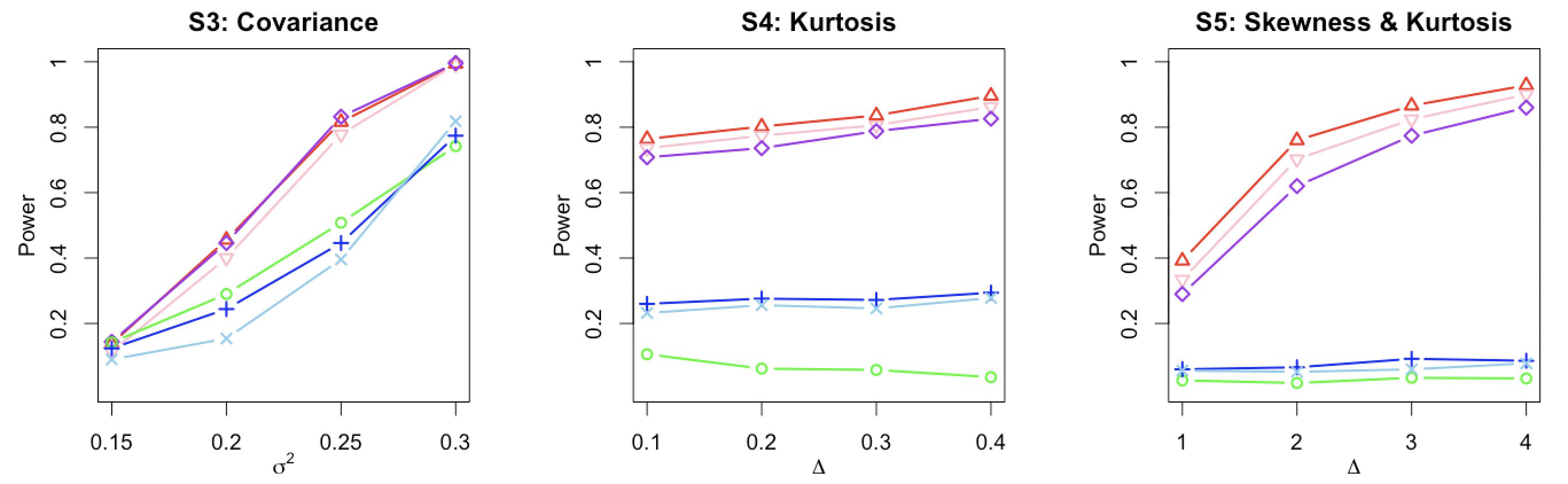}
	\includegraphics[width=\columnwidth]{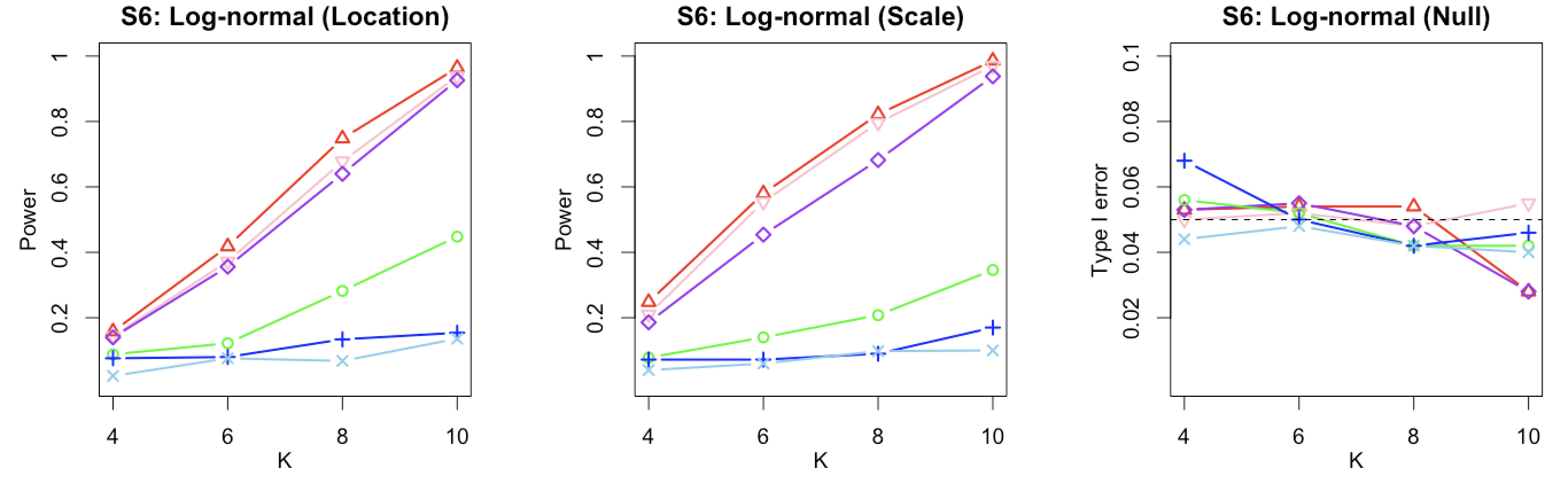}
	\includegraphics[width=\columnwidth]{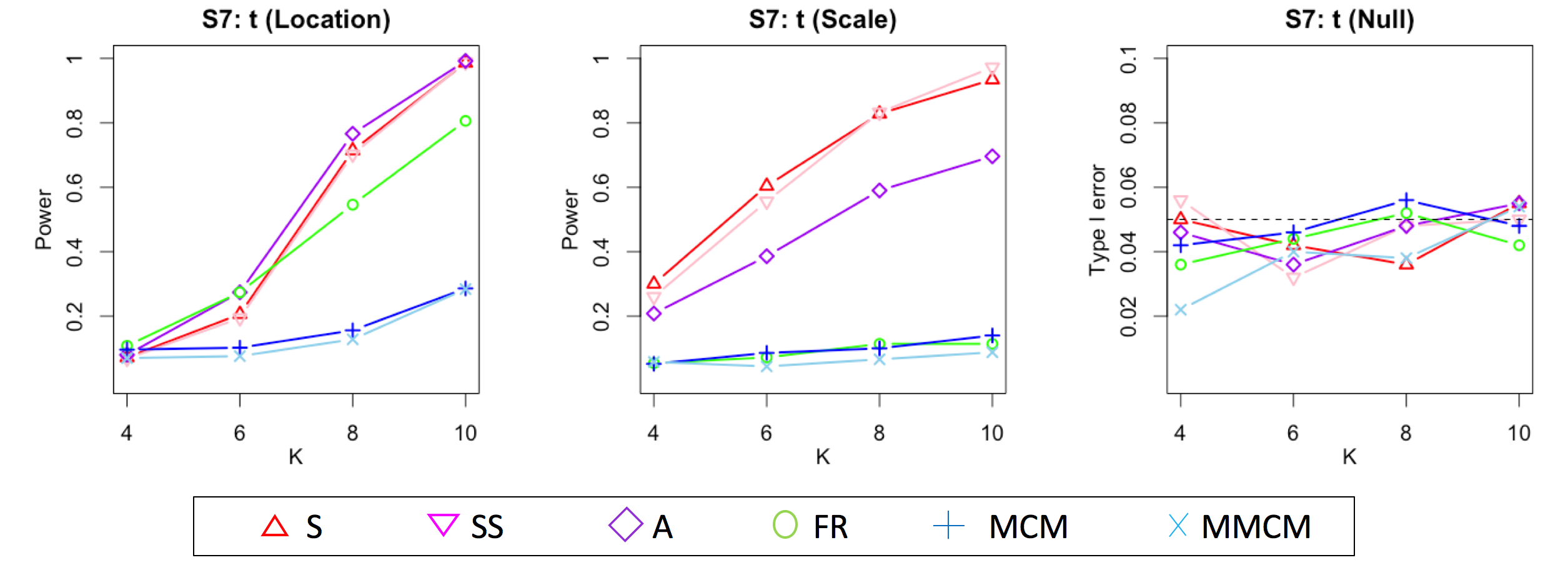}
	\caption{Simultation results under different scenarios S1--S7.} 
	\label{fig:dim}
\end{figure}

The results are shown in Figure \ref{fig:dim}. For location alternatives (S1), we see that the new tests in general outperform other tests, especially for high-dimensional cases. In particular, $S$ and $A$ exhibit high power, followed by $SS$. $FR$ also does well, but it is outperformed by the new tests as the dimension or $K$ increases. For scale alternatives (S2), the existing tests are drastically outperformed by the new tests. Among the new tests, $S$ shows the best performance and followed by $SS$, and then by $A$.  For  covariance differences (S3), we see that the new tests dominate in power. Moreover, the results under the scenarios S4 and S5 show that the new approaches are also very sensitive to differences in the skewness and kurtosis, while the existing tests cannot capture these differences. The results under scenarios S6 and S7 show that the new tests in general outperform the existing tests for the multivariate log-normal and $t$-distributed data, and the new tests work well for both symmetric and asymmetric distributions under moderate to high dimensions. The new tests also control the type I error well.

The overall pattern of the simulation results shows that the new tests improve with increasing separation and dimension and show high power for a wide range of alternatives. In particular, $S$ and $A$ exhibit high power for location alternatives and $S$ and $SS$ exhibit high power for scale alternatives. In practice, $SS$ would be preferred over $S$ as it is fast and effective to general alternatives. If further investigation is needed, the permutation test based on $S$ would also be useful.



\section{A real data example} \label{sec:real}

We illustrate the new tests on the age dataset, namely the IMDb-WIKI database\footnote{https://data.vision.ee.ethz.ch/cvl/rrothe/imdb-wiki/} \cite{rothe2018deep}. The age dataset consists of 397,949 images of 19,545 celebrities with corresponding age labels.  The data contain no personally identifiable information. Here, we follow the preprocessing of \cite{law2018bayesian} where they use the representation $\phi(x): \mathcal{R}^{256\times256}\rightarrow\mathcal{R}^{4096}$, mapping from the pixel space of images to the CNN's last hidden layer learnt by \cite{rothe2018deep}. We construct five groups $(K=5)$ according to the celebrity's age label, 10-20, 20-30, 30-40, 40-50, and 50-60, where for example 10-20 indicates the images corresponding to the celebrity's age label that is between 10 and 20.

We utilize the age dataset to examine how the new tests distinguish the images depending on the celebrity's age. To this end, we conduct the testing procedures on subsets of the whole data so that we can approximate the empirical power of the tests. We simulate 1,000 randomly selected subsets of the data from each age group with sample sizes $n_{i}$ $(i=1,\ldots,5)$. Here, the significance level is set to be 0.01 for all tests.

The first table of Table \ref{tab:age} shows the estimated power of the tests under different sample sizes. We see that the power of the tests increases as the sample size increases and the new tests outperform the existing tests in all cases. We further check the pattern of statistics under different cases (the second table of Table \ref{tab:age}). We simulate 10,000 datasets when $n_{i} = 10$ and there are 1,493 cases where all tests reject the null (Case 1), 15 cases where the only existing tests reject the null (FR, MCM, MCMM all reject; none of S, SS, A reject) (Case 2), and 558 cases where the only new tests reject the null (S, SS, A all reject; none of FR, MCM, MCMM reject) (Case 3) at 0.01 significance level. 

\begin{figure}[h]
	\centering
	\includegraphics[width=\textwidth]{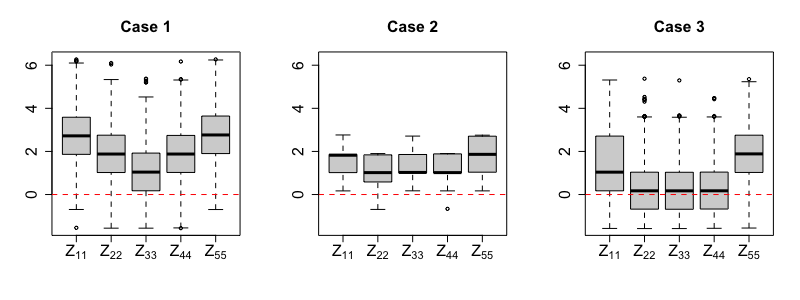} 
	\begin{tabular}{ccccccc}
	\toprule
	& S & SS & A & FR & MCM & MCMM \\ \cline{2-7}
	$p$-value & 0.018 & 0.042 & 0.098 & 0.005 & 0.005 & 0.000 \\
	\bottomrule
	\end{tabular}
	\caption{\label{fig:boxplot} Boxplots of the within-sample statistics under different cases and average $p$-values of the tests in Case 2.} 
\end{figure}

\begin{wraptable}{r}{0.6\linewidth}
	\centering
	\caption{Estimated power of the tests for the age dataset and the number of rejecting the null under different cases\label{tab:age}}
	\begin{tabular}{cccccc}
		\toprule
		$n_{i}$ & 5 & 10 & 15 & 20 & 25 \\
		\midrule
		S &  \textbf{0.365} & \textbf{0.685} &  \textbf{0.880} & \textbf{0.956} &  \textbf{0.987}  \\
		SS &  0.269 & 0.601 &  0.812& 0.937 &  0.976  \\
		A &  0.271 & 0.563 &  0.796 & 0.916 &  0.974  \\
		FR &  0.250 & 0.598 &  0.802 & 0.925 &  0.968  \\
		MCM  &  0.166 &   0.437 &   0.655 & 0.777 & 0.881 \\
		MMCM & 0.069 & 0.215 & 0.425 & 0.591 & 0.766 \\
		\bottomrule
	\end{tabular}

    \vspace*{0.3cm}
    
    \begin{tabular}{cccc}
    	\toprule
    	Trials & Case 1 & Case 2 & Case 3  \\
    	10,000 &  1,493 & 15 & 558  \\
    	\bottomrule
    \end{tabular}
\end{wraptable}
We take a closer look at the test statistics and Figure \ref{fig:boxplot} shows boxplots of within-sample statistics under each case, where $Z_{ii} = \left(R_{ii}-\ep(R_{ii})\right)/\sqrt{\var(R_{ii})}$ $(i=1,\ldots,5)$. Here, relatively large values of $Z_{ii}$ are the evidence against the null. In Case 3, $Z_{22}$, $Z_{33}$, and $Z_{44}$ are close to zero, which thus leads to poor performance of the existing tests. The new tests take into account both within-sample and between-sample statistics, and are more powerful. On the other hand, since the new tests are developed to cover a wide range of alternatives, the existing tests aimed at specific alternatives, e.g., location alternatives with small $K$ or symmetrically distributed alternatives, could show better performance. On other hand, even in Case 2, the new tests exhibit relatively small $p$-values as well (Table in Figure \ref{fig:boxplot}). More data analysis can be found in Appendix E.
\vspace{-1em}


\section{Discussion} \label{sec:discussion}

In this paper, we propose new graph-based multi-sample tests for comparing multiple distributions by utilizing information embedded in groups as much as possible. When the number of samples or groups is large, computing pairwise distances for constructing the similarity graph could be computationally expensive. In this case, faster algorithms, such an approximate nearest neighbor algorithm that do not compute all pairwise distances \cite{beygelzimer2013fnn}, could be used to save time.




\bibliographystyle{plain}
\bibliography{references}

\section*{Checklist}


\begin{enumerate}

\item For all authors...
\begin{enumerate}
  \item Do the main claims made in the abstract and introduction accurately reflect the paper's contributions and scope?
    \answerYes{}
  \item Did you describe the limitations of your work?
    \answerYes{See Section \ref{sec:real}.}
  \item Did you discuss any potential negative societal impacts of your work?
    \answerNo{}
  \item Have you read the ethics review guidelines and ensured that your paper conforms to them?
    \answerYes{}
\end{enumerate}

\item If you are including theoretical results...
\begin{enumerate}
  \item Did you state the full set of assumptions of all theoretical results?
    \answerYes{See Theorem \ref{thm:asymp} and Theorem \ref{thm:consistency}.}
        \item Did you include complete proofs of all theoretical results?
    \answerYes{See Appendix A and C.}
\end{enumerate}

\item If you ran experiments...
\begin{enumerate}
  \item Did you include the code, data, and instructions needed to reproduce the main experimental results (either in the supplemental material or as a URL)?
    \answerYes{See the “code” folder in the supplementary material.}
  \item Did you specify all the training details (e.g., data splits, hyperparameters, how they were chosen)?
    \answerYes{See Section \ref{sec:numerical}.}
        \item Did you report error bars (e.g., with respect to the random seed after running experiments multiple times)?
    \answerNo{}
        \item Did you include the total amount of compute and the type of resources used (e.g., type of GPUs, internal cluster, or cloud provider)?
    \answerNo{}
\end{enumerate}

\item If you are using existing assets (e.g., code, data, models) or curating/releasing new assets...
\begin{enumerate}
  \item If your work uses existing assets, did you cite the creators?
    \answerYes{See Section \ref{sec:real} for age image dataset.}
  \item Did you mention the license of the assets?
    \answerYes{See Section \ref{sec:real}.}
  \item Did you include any new assets either in the supplemental material or as a URL?
    \answerYes{See  Section \ref{sec:real} and “code” folder in the supplementary material.}
  \item Did you discuss whether and how consent was obtained from people whose data you're using/curating?
    \answerYes{See Section \ref{sec:real}.}
  \item Did you discuss whether the data you are using/curating contains personally identifiable information or offensive content?
    \answerYes{See Section \ref{sec:real}.}
\end{enumerate}

\item If you used crowdsourcing or conducted research with human subjects...
\begin{enumerate}
  \item Did you include the full text of instructions given to participants and screenshots, if applicable?
    \answerNA{We did not conduct research with human subjects.}
  \item Did you describe any potential participant risks, with links to Institutional Review Board (IRB) approvals, if applicable?
    \answerNA{}
  \item Did you include the estimated hourly wage paid to participants and the total amount spent on participant compensation?
    \answerNA{}
\end{enumerate}

\end{enumerate}


\end{document}